# Lattice-Driven Magnetoresistivity and Metal-Insulator Transition in Single-Layered Iridates

M. Ge[1,2,3], T. F. Qi[1,2], O. B. Korneta[1,2], D. E. De Long[1,2], P. Schlottmann[4], W. P. Crummett[1,5] and G. Cao[1,2*]

[1]Center for Advanced Materials, University of Kentucky

[2]Department of Physics and Astronomy, University of Kentucky

[3]China High Magnetic Field Lab and University of Science and Technology of China

[4]Department of Physics, Florida State University

[5]Department of Physics, Centre College

$Sr_2IrO_4$ exhibits a novel insulating state driven by spin-orbit interactions. We report two novel phenomena, namely a large magnetoresistivity in $Sr_2IrO_4$ that is extremely sensitive to the orientation of magnetic field but exhibits no apparent correlation with the magnetization, and a robust metallic state that is induced by dilute electron ($La^{3+}$) or hole ($K^+$) doping for $Sr^{2+}$ ions in $Sr_2IrO_4$. Our structural, transport and magnetic data reveal that a strong spin-orbit interaction alters the balance between the competing energies so profoundly that (1) the spin degree of freedom alone is no longer a dominant force; (2) underlying transport properties delicately hinge on the Ir-O-Ir bond angle via a strong magnetoelastic coupling; and (3) a highly insulating state in $Sr_2IrO_4$ is proximate to a metallic state, and the transition is governed by lattice distortions. This work suggests that a novel class of lattice-driven electronic materials can be developed for applications.

The 5d-based iridates have become fertile ground for studies of new physics driven by strong spin-orbit interaction; this new physics is embodied by a large array of novel phenomena observed recently, such as $J_{eff}$ = 1/2 Mott insulator [1-3], hyper-kagome structure [4], giant magnetoelectric effect [5], exotic metallic states [6,7], unusual orbital magnetism [8], etc. The list of relevant theoretical proposals is already long and intriguing: high $T_c$ superconductivity [9], correlated topological insulator [10, 11], Dirac semimetal with Fermi arcs [12], Kitaev mode [12, 13], etc. It is known that relativistic spin-orbit interaction proportional to $Z^4$ (Z the atomic number) ranges from 0.2 to 1 eV in 5d materials (compared to ~ 20 meV in 3d materials), therefore it can no longer be treated as a perturbation as is in many other materials where the magnetic interaction is dominated by the spin degree of freedom alone. Instead, the strong spin-orbit interaction vigorously competes with Coulomb (0.5 – 2 eV) and other interactions, thus sets a new balance between the relevant energies that drives exotic states seldom or never seen in other materials; findings reported here constitute a compelling example.

$Sr_2IrO_4$, with a crystal structure similar to that of $La_2CuO_4$ and the p-wave superconductor $Sr_2RuO_4$, is a weak ferromagnet (FM) with a Curie temperature $T_C$ = 240 K [14-16]. A unique and important structural feature of $Sr_2IrO_4$ is that it crystallizes in a reduced tetragonal structure (space-group $I4_1/acd$) due to a rotation of the $IrO_6$-octahedra about the c-axis by ~11°, resulting in a larger unit cell by $\sqrt{2}$ x $\sqrt{2}$ x 2 [14, 15]. This rotation corresponds to a distorted in-plane Ir1-O2-Ir1 bond angle θ critical to the electronic structure [7, 9, 12, 20]. It is already established that $Sr_2IrO_4$ is a novel Mott insulator dictated by spin-orbit interaction [1-3]. In essence, strong crystal fields split off 5d band states with $e_g$ symmetry, and $t_{2g}$ bands arise from $J_{eff}$ = 1/2 and $J_{eff}$ = 3/2 multiplets via strong spin-orbit interaction (~0.4 eV). A weak admixture of the $e_g$ orbitals downshifts the $J_{eff}$ = 3/2 quadruplet from the $J_{eff}$ = 1/2 doublet. Since the $Ir^{4+}$ ($5d^5$) ions provide five electrons, four of them fill the lower $J_{eff}$ = 3/2 bands, and one electron partially fills the $J_{eff}$ = 1/2 band. The $J_{eff}$ = 1/2 band is so narrow that even a reduced coulomb repulsion U (~ 0.5 eV) is sufficient to open a small gap supporting the insulating state [1, 2]. A similar mechanism also describes insulating states observed in other iridates, such as $Sr_3Ir_2O_7$ [2, 17] and $BaIrO_3$ [8, 18].

In this Letter, we report the following central findings for single-crystal $Sr_2IrO_4$ and its derivatives with dilute doping: (1) the magnetic structure varies with temperature T, resulting in three temperature regions that show distinct magneto-transport behavior; (2) the isothermal resistivity ρ(H) exhibits a large, lattice-driven magnetoresistivity punctuated with multiple transitions that is highly sensitive to the orientation of the magnetic field H, but shows no apparent correlation with the isothermal magnetization M(H) when H ∥ c-axis; and (3) a robust metallic state is readily induced by dilute doping of either $La^{3+}$ or $K^+$ ions for $Sr^{2+}$ ions in $Sr_2IrO_4$, highlighting a proximity of the insulating state to a metallic state that is mainly controlled by the lattice degrees of freedom. The spin degree of freedom alone is no longer a driving force due to the strong spin-orbit interaction. These novel phenomena open a new avenue for studies of physics driven by spin-orbit coupling, and also pose new device paradigms for lattice-driven electronic materials.

Single crystals studied were synthesized using a self-flux technique described elsewhere [5-7, 16-18]. The average size of the single crystals is 1.0 x 0.7 x 0.2 $cm^3$. The structures of $(Sr_{1-x}La_x)_2IrO_4$ and $(Sr_{1-x}K_x)_2IrO_4$ were determined using a Nonius Kappa CCD X-Ray Diffractometer with sample temperature controlled using a nitrogen stream. Structures were refined by full-matrix least-squares using the SHELX-97 programs [21]. Chemical compositions of the single crystals were determined using energy dispersive X-ray analysis (EDX). Resistivity

ρ(T,H) and magnetization M(T,H) were measured using a Quantum Design (QD) 7T SQUID Magnetometer and a QD 14T Physical Property Measurement System, respectively.

This study captures a few critical magnetic features of $Sr_2IrO_4$ that need to be addressed first. While both the a-axis $M_a(T)$ and the c-axis $M_c(T)$ expectedly show ferromagnetic (FM) order below $T_C$ = 240 K, and a positive Curie-Weiss temperature, $\theta_{cw}$ = +236 K, confirm the FM exchange coupling at high T [5, 7, 14-16], a close examination of the low-field M(T) reveals *two additional anomalies* at $T_{M1} \approx$ 100 K and $T_{M2} \approx$ 25 K in $M_a(T)$ and $M_c(T)$ (see Fig.1a). Our previous ac magnetic susceptibility also exhibits a peak near $T_{M1}$ as well as frequency dependence indicative of magnetic frustration [5]. Indeed, a recent muon-spin rotation (μSR) study of $Sr_2IrO_4$ reports two structurally equivalent muon sites that experience increasingly distinct local magnetic fields for T < 100 K, which subsequently lock in below 20 K [19]. It becomes clear that the magnetic structure varies with T, resulting in *three well-defined temperature Regions I, II, and III* (Fig.1a), which exhibit distinct physical properties, as discussed below. Moreover, $M_a(T)$ decreases rapidly below $T_{M1}$ and $T_{M2}$, but $M_c(T)$ rises below 50 K and more sharply below $T_{M2}$ as T decreases (see Fig.1a inset). The different T-dependences of $M_a(T)$ and $M_c(T)$ signal an evolving magnetic structure where *the spins may no longer lie within the basal plane below $T_{M1}$*. This spin reorientation apparently simultaneously weakens $M_a$ but enhances $M_c$ thereby reducing the magnetic anisotropy $M_a/M_c$, which decreases from 2.2 at 100 K to 1.5 at 1.7 K (see Figs.1b and 1c).

The electrical resistivity for the a-axis $\rho_a$ (T) follows an activation law, $\rho_a$ (T) ~ exp ($\Delta/2k_BT$) (where $\Delta$ is the energy gap and $k_B$ the Boltzmann's constant), and exhibits three distinct values of $\Delta$ in regions that closely correspond to Regions I, II and III defined above, as shown in Fig.1d. It is noteworthy that $\Delta$ in Region III is quite close to the optically measured gap (~ 0.1 eV) [1];

and it further narrows with decreasing T (Fig.1d) and, unexpectedly, with the application of a modest magnetic field of a few Tesla (not shown).

Indeed, the transport properties are coupled to H in such a peculiar fashion that no current models can describe the observed magnetoresistivity shown in Figs. 2 and 3. We focus on a representative temperature T = 35 K that is within Region II. For H ∥ a-axis, both the a-axis resistivity $\rho_a$(H∥a) (Fig.2b) and the c-axis resistivity $\rho_c$(H∥a) (Fig.2c) exhibit an abrupt drop by ~ 60% near $\mu_o$H = 0.4 T. These data only partially track the field dependences of $M_a$(H) and $M_c$(H) shown in Fig. 2a; and given the ordered moment $m_s$ < 0.07 $\mu_B$/Ir, the reduction of spin scattering alone certainly cannot account for such a drastic reduction in $\rho$(H). Even more strikingly, for H ∥ c-axis, *both $\rho_a$(H∥c) and $\rho_c$(H∥c) exhibit multiple anomalies at $\mu_o$H = 2 T and 3 T, which leads to a large overall resistivity reduction by more than 50%; however, no anomalies corresponding to these transitions in $M_a$(H) and $M_c$(H) are discerned*! Such behavior is clearly not due to the Lorenz force because $\rho_c$(H∥c) exhibits the same behavior in a configuration where both the current and H are parallel to the c-axis (Fig.2c); the conspicuous lack of the correlation between $\rho$ and M is apparently not endorsed by any existing models describing magnetoresistivity observed in other known materials.

An essential contributor to conventional magnetoresistance is spin-dependent scattering; negative magnetoresistance can be a result of the reduction of spin scattering due to spin alignment with increasing magnetic field. The data in Fig. 2 therefore raise a fundamental question: *Why does the resistivity sensitively depend on the orientation of magnetic field H but show no direct relevance to the measured magnetization when H∥c-axis?* While no conclusive answers to the question are yet available, one scenario may be qualitatively relevant.

This scenario is based on the following understanding established in this and previous work of $Sr_2IrO_4$: (1) In the case of strong spin-orbit interaction, the lattice distortion, or specifically, the Ir1-O2-Ir1 bond angle θ dictates the low-energy Hamiltonian [12], and the band structure [7, 20]. (2) A strong spin-orbit interaction can cause the spins to rigidly rotate with the $IrO_6$-octahedra via strong spin-lattice or magnetoelastic coupling [5, 12]. (3) The reduced magnetic anisotropy $M_a/M_c$ strongly indicates an emerging c-axis spin component below $T_{M1}$ that generates a noncollinear spin structure and frustration, as manifested in Fig.1, and in previous studies [5, 12, 19]; the noncollinearity could take the form of a spiral spin configuration where the spin direction is rigidly maintained at an angle β with respect to the c-axis, as sketched in Fig. 2d.

Recent studies of $Sr_2IrO_4$ have already established that electron hopping sensitively depends on the bond angle θ [7]. In particular, hopping occurs through two active $t_{2g}$ orbitals: $d_{xy}$ and $d_{xz}$ for θ = 180°, and $d_{xz}$ and $d_{yz}$ for θ = 90° [12]. It is recognized that the larger θ, the more energetically favorable it is for electron hopping and superexchange interactions. Since the $IrO_6$-octahdra rotate with the spins, the application of H ∥ c-axis must at least slightly rotate the $IrO_6$-octahdra about the c-axis, which, in turn, changes θ. It is important to realize that even a small increase in θ due to increasing H can be sufficient to drastically enhance the hopping, which could explain the multiple downturns in ρ(H). The clear hysteresis exhibited in Fig.2b reinforces the notion that the magnetoresistivity is primarily driven by field-induced lattice distortions when H ∥ c. The absence of anomalies in $M_a(H)$ and $M_c(H)$ corresponding to the transitions in $ρ_a(H)$ and $ρ_c(H)$ can be attributed to a spiral spin configuration: the spins respond to H only by rotating about the c-axis, and this rotation changes θ but does not affect β or the c-axis projection of the magnetic moment, as schematically illustrated in Fig.2d; therefore, $M_a(H)$ and $M_c(H)$ remain unchanged.

The delicate nature of the coupling of the magneto-transport behavior to the lattice and magnetic structure is apparent in a few respects, as illustrated in Fig. 3. The transport behavior seen in Region II is no longer observable in Region I, where the magnetoresistivity is remarkably weak; this is evident in $\rho_a(H)$ and $\rho_c(H)$ at T = 10 K, as shown in Fig. 3a. Moreover, the application of *H||a-axis* causes a pronounced rise in $\rho_a(H)$ rather than the sharp drop observed in Region II at low H (Figs. 2, 3b and 3c), and a reversal of the resistivity anisotropy (Fig. 3a). On the other hand, as T approaches Region III, the field dependence of $\rho_a(H)$ and $\rho_c(H)$ retains some resemblance to that in the Region II, but becomes far weaker.

Indeed, the ground state can be readily changed via slight manipulations of θ. As documented in Fig. 4a, a dilute doping of either $La^{3+}$ or $K^+$ ions for $Sr^{2+}$ ions leads to a larger θ despite considerable differences between the ionic radii of Sr, La, and K, which are 1.18 Å, 1.03 Å and 1.38 Å, respectively. Since hopping between active $t_{2g}$ orbitals is critically linked to θ, drastic changes in physical properties due to such sizable increases in θ are anticipated. It is therefore understandable that $\rho_a$ ($\rho_c$) is reduced by a factor of *$10^8$* (*$10^{10}$*) at low T as x evolves from 0 to 0.04 and 0.02 for La and K, respectively (see Figs. 4b, 4c and 4d). For a La doping of x = 0.04, there is a sharp downturn near 10 K, indicative of a rapid decrease in inelastic scattering (Figs. 4c inset). Such low-T behavior is similar to that observed in slightly oxygen depleted $Sr_2IrO_{4-\delta}$ with δ = 0.04 [7]. It is noteworthy that $T_C$ decreases with La doping in $(Sr_{1-x}La_x)_2IrO_4$ (not shown), and vanishes at x=0.04 where the metallic state is fully established; in contrast, the magnetically ordered state coexists with the fully metallic state in $(Sr_{0.98}K_{0.02})_2IrO_4$, as shown in Fig. 4d. This comparison stresses that the occurrence of a metallic state does not necessarily accompany radical changes in the magnetic state in iridates; this observation is in accord with a conspicuous characteristic of $Sr_2IrO_4$ where the resistivity shows no anomaly near $T_C$ (= 240 K) [5,7,16]. The radical changes in transport properties of $Sr_2IrO_4$ with dilute doping strongly suggest that the insulating state driven by a strong spin-orbit interaction is proximate to a metallic state. *The inducement of a robust metallic state by either dilute electron ($La^{3+}$) or hole ($K^+$) doping for $Sr^{2+}$ further reinforces the central finding of this work that transport properties in iridates such as $Sr_2IrO_4$ can be chiefly dictated by the lattice degrees of freedom.*

In summary, the large and uniquely anisotropic magnetoresistivity in $Sr_2IrO_4$ and the robust metallic state in doped $Sr_2IrO_4$ are attributed to the subtle unbuckling of the $IrO_6$-octahedra, without apparent correlation with the magnetization as conventionally anticipated. We conclude that a strong spin-orbit interaction fundamentally changes the balance between the competing energies such that (1) the spin degree of freedom alone is no longer a dominant variable; (2) electron hopping delicately depends upon the lattice distortion via strong magnetoelastic coupling; and (3) the highly insulating state in $Sr_2IrO_4$ is proximate to a metallic state. We expect such a novel magnetotransport behavior to result in new technological paradigms based upon "lattice-driven electronic materials"

One of us (GC) is very thankful to Profs G. Murphy and R. Kaul for enlightening discussions. This work was supported by NSF through grants DMR-0856234 (GC) and EPS-0814194 (GC, LED), and by DoE through grants DE-FG02-97ER45653 (LED) and DE-FG02-98ER45707 (PS).

* Corresponding author; email: cao@uky.edu

References

1. B. J. Kim, et al, *Phys. Rev. Lett.* 101, 076402 (2008)

2. S. J. Moon, et al, *Phys. Rev. Lett.* 101, 226401 (2008)

3. B.J. Kim, et al, *Science* 323, 1329 (2009)

4. Y. Okamoto, et al, *Phys. Rev. Lett.* 99, 137207 (2007)

5. S. Chikara, et al, *Phys. Rev. B* 80, 140407 (R) (2009)

6. G. Cao, et al, *Phys. Rev. B* 76, 100402(R) (2007)

7. O.B. Korneta, et al, *Phys. Rev. B* 82 115117 (2010)

8. M. A. Laguna-Marco, et al, *Phys. Rev. Lett.* 106, 136402 (2011)

9. Fa Wang and T. Senthil, *Phys. Rev. Lett.* 106, 136402 (2011)

10. A. Shitade, et al, *Phys. Rev. Lett.* 102, 256403 (2009)

11. D.A. Pesin, and Leon Balents, *Nat. Phys.* 6, 376 (2010)

12. G. Jackeli and G. Khaliulin, *Phys. Rev. Lett.* 102 017205 (2009)

13. Xiangang Wan, et al, *Phys. Rev. B* 83, 205101 (2011)

14. Q. Huang, et al, *J. Solid State Chem.* 112, 355 (1994)

15. R.J. Cava, et al, *Phys. Rev. B* 49, 11890 (1994)

16. G. Cao, et al, *Phys. Rev. B* 57, R11039 (1998)

17. G. Cao, et al, *Phys. Rev. B* 66, 214412 (2002)

18. G. Cao, et al, *Solid State Commun.* 113, 657 (2000)

19. I. Franke, et al, *Phys. Rev. B* 83 094416 (2011)

20. S. J. Moon, et al, *Phys. Rev. B* 80, 195110 (2009)

21. G. M. Sheldrick, *Acta Crystallogr* A 64, 112 (2008)

Captions

Fig.1. The magnetization for the a-axis and the c-axis, $M_a$ and $M_c$, as a function of (a) temperature at $\mu_oH=0.1$ T, and (b) magnetic field at T=1.7 and 100 K; (c) The magnetic anisotropy $M_a/M_c$ as a function of temperature; (d) The resistivity for the a-axis ln $\rho_a$ as a function of 1/T; Inset in (a): Enlarged low-T $M_c$. Note that the data in (a) and (d) define Regions I, II, and III.

Fig.2. The field dependence at T = 35 K of (a) the magnetization $M_a$ and $M_c$; (b) The a-axis resistivity $\rho_a$ for H||a and H||c; and (c) The c-axis resistivity $\rho_c$ for H||a and H||c; (d) The schematics of the spin configuration for the basal plane (left) and the ac-plane (right).

Fig.3. The field dependence of the a-axis resistivity $\rho_a$ for H||a and H||c at representative temperatures at (a) T = 10 K (Region I), (b) T = 30 K, (c) T = 50 K, and (d) T = 100 K (approaching Region III).

Fig.4. (a) The Ir1-O2-Ir1 bond angle $\theta$ as a function of La and K doping concentration x. The temperature dependence of (b) the a-axis resistivity $\rho_a$, and (c) the c-axis resistivity $\rho_c$ for $(Sr_{1-x}La_x)_2IrO_4$ with $0 \leq x \leq 0.04$; (d) The temperature dependence of $\rho_a$ and $M_a$ at $\mu_oH=0.1$ T (right scale) for $(Sr_{0.98}K_{0.02})_2IrO_4$. Inset in (c): Enlarged low-T $\rho_c$.

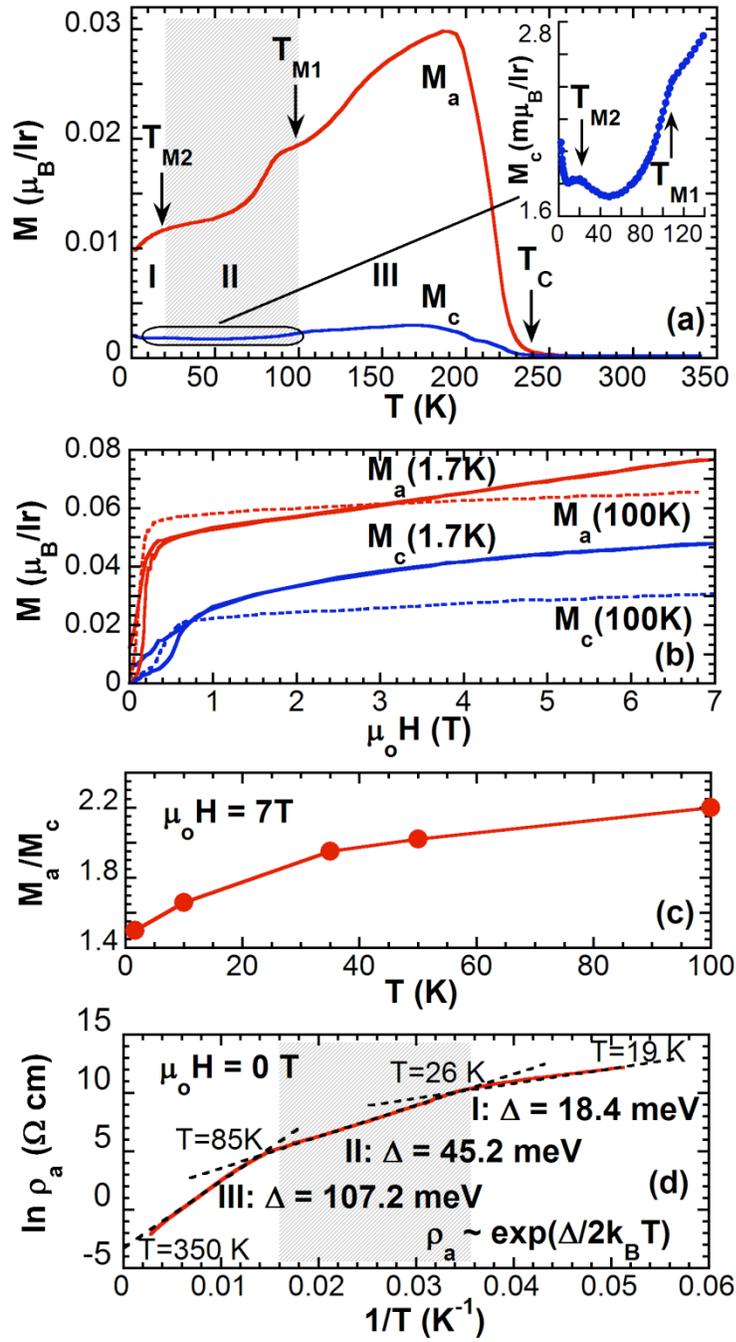

Fig.1

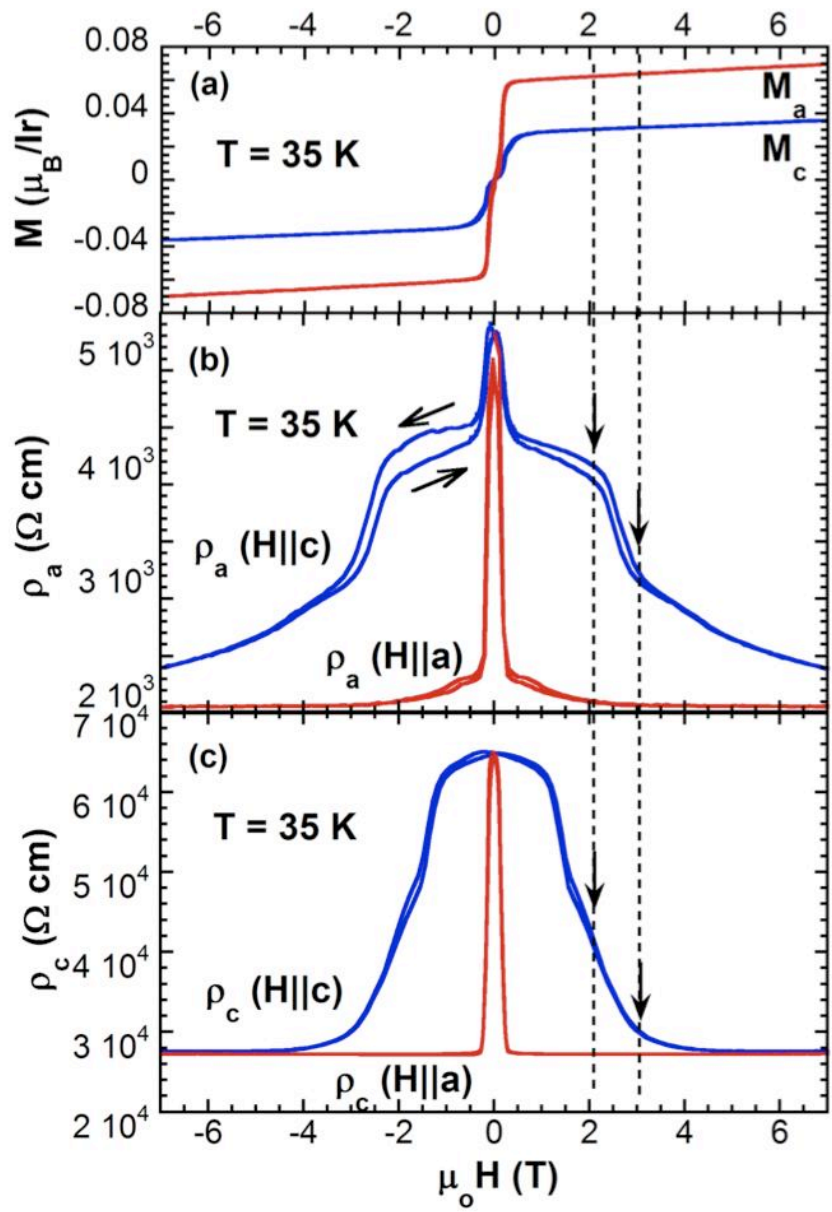

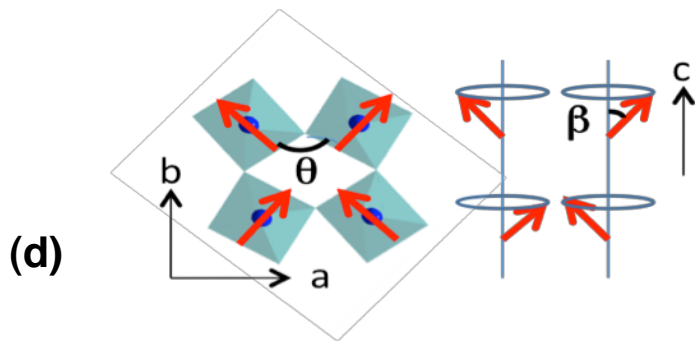

Fig. 2

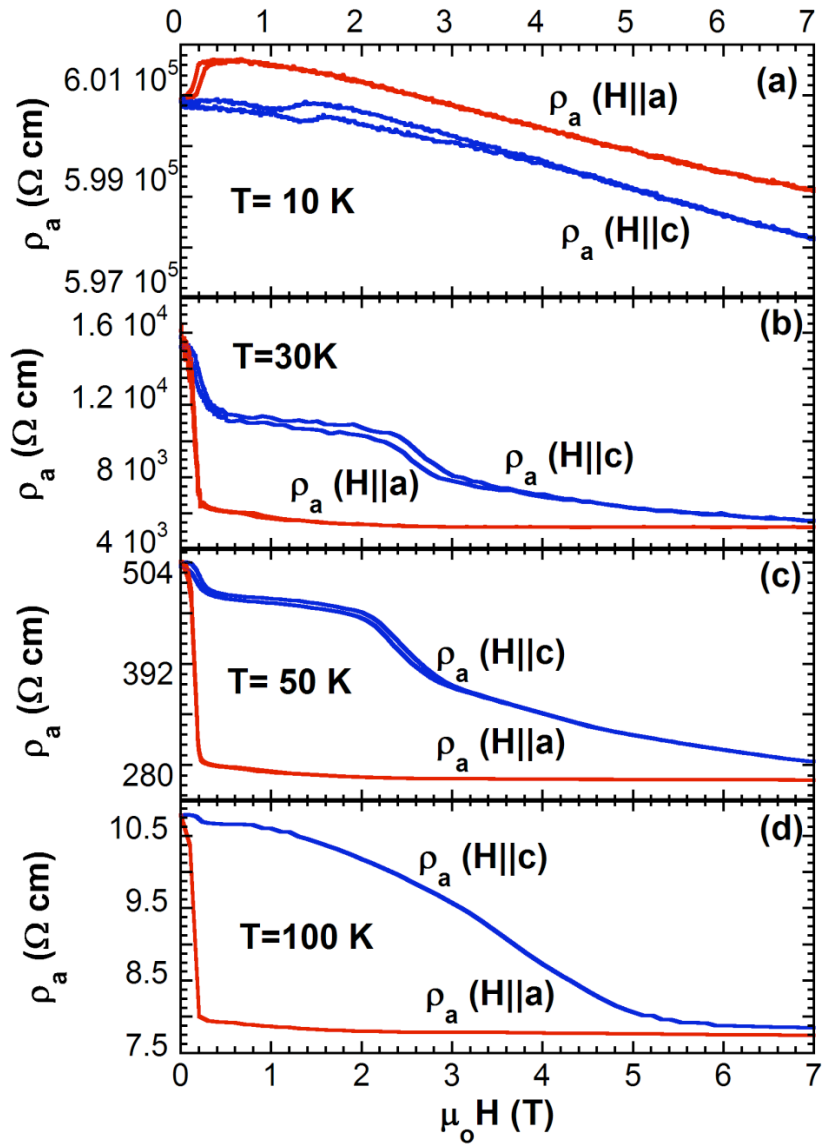

Fig.3

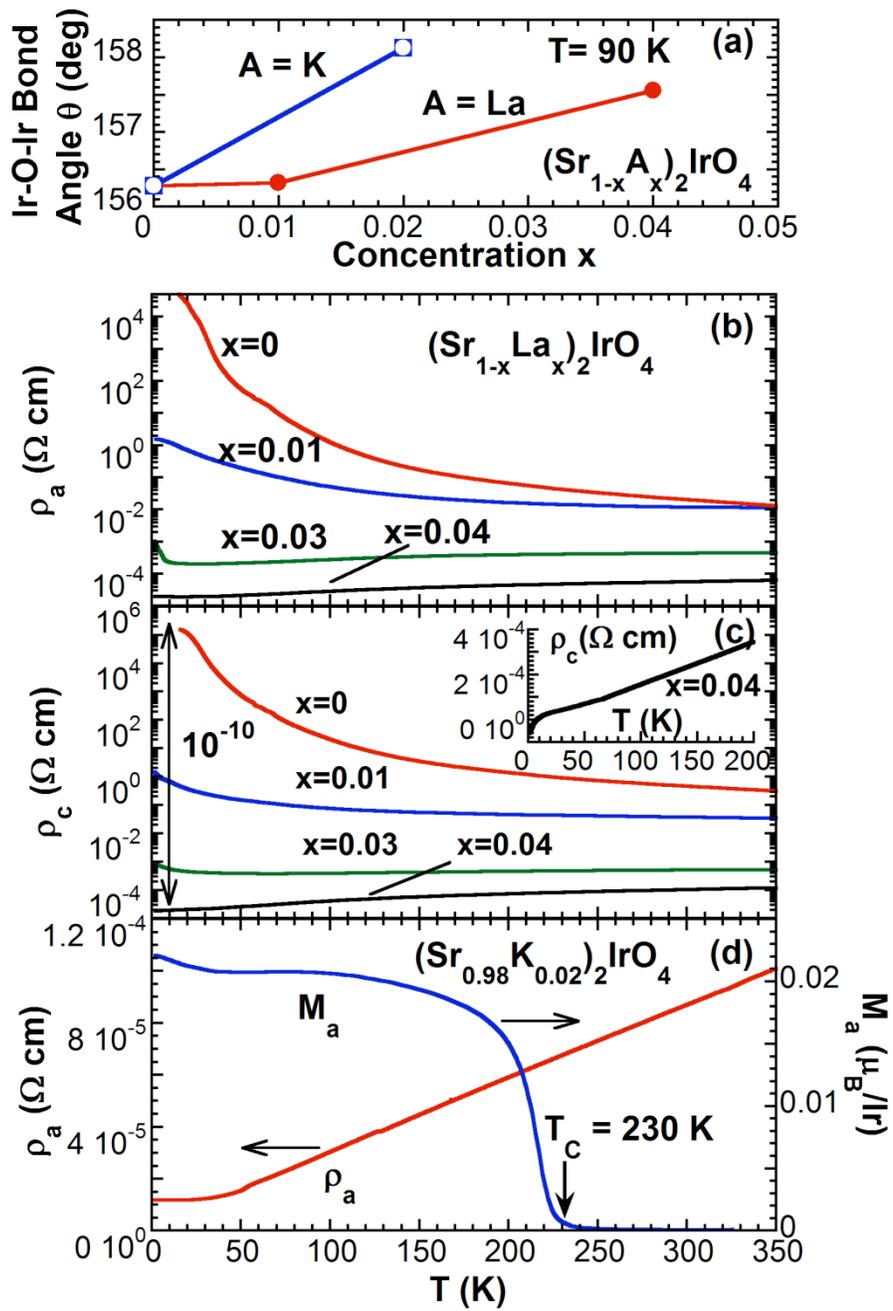

Fig.4
14